\documentclass[sigconf,authorversion,nonacm]{acmart}

\usepackage{balance}
\usepackage{caption}
\usepackage{subcaption}

\AtBeginDocument{%
  \providecommand\BibTeX{{%
    \normalfont B\kern-0.5em{\scshape i\kern-0.25em b}\kern-0.8em\TeX}}}

\copyrightyear{2021}
\acmYear{2021}
\setcopyright{acmlicensed}
\acmConference[ICMI '21]{Proceedings of the 2021 International Conference on Multimodal Interaction}{October 18--22, 2021}{Montréal, QC, Canada}
\acmBooktitle{Proceedings of the 2021 International Conference on Multimodal Interaction (ICMI '21), October 18--22, 2021, Montréal, QC, Canada}
\acmPrice{15.00}
\acmISBN{978-1-4503-8481-0/21/10}
\acmDOI{10.1145/3462244.3479910}

\begin{document}
\fancyhead{}

\title[ML-PersRef: A Machine Learning-based Personalized Multimodal Fusion Approach for Referencing]{ML-PersRef: A Machine Learning-based Personalized Multimodal Fusion Approach for Referencing Outside Objects From a Moving Vehicle}


\author{Amr Gomaa}
\affiliation{%
  \institution{German Research Center for Artificial Intelligence (DFKI)}
  \city{Saarbr{\"u}cken}
  \country{Germany}
}
\email{amr.gomaa@dfki.de}

\author{Guillermo Reyes}
\affiliation{%
  \institution{German Research Center for Artificial Intelligence (DFKI)}
  \city{Saarbr{\"u}cken}
  \country{Germany}
}
\email{guillermo.reyes@dfki.de}

\author{Michael Feld}
\affiliation{%
  \institution{German Research Center for Artificial Intelligence (DFKI)}
  \city{Saarbr{\"u}cken}
  \country{Germany}
}
\email{michael.feld@dfki.de}


\begin{abstract}

Over the past decades, the addition of hundreds of sensors to modern vehicles has led to an exponential increase in their capabilities. This allows for novel approaches to interaction with the vehicle that go beyond traditional touch-based and voice command approaches, such as emotion recognition, head rotation, eye gaze, and pointing gestures. Although gaze and pointing gestures have been used before for referencing objects inside and outside vehicles, the multimodal interaction and fusion of these gestures have so far not been extensively studied. 
We propose a novel learning-based multimodal fusion approach for referencing outside-the-vehicle objects while maintaining a long driving route in a simulated environment. The proposed multimodal approaches outperform single-modality approaches in multiple aspects and conditions. Moreover, we also demonstrate possible ways to exploit behavioral differences between users when completing the referencing task to realize an adaptable personalized system for each driver. We propose a personalization technique based on the transfer-of-learning concept for exceedingly small data sizes to enhance prediction and adapt to individualistic referencing behavior. Our code is publicly available at \url{https://github.com/amr-gomaa/ML-PersRef}.

\end{abstract}

\ccsdesc[500]{Human-centered computing~HCI design and evaluation methods}
\ccsdesc[500]{Human-centered computing~Pointing}
\ccsdesc[500]{Human-centered computing~Gestural input}
\ccsdesc[300]{Computing methodologies~Support vector machines}
\ccsdesc[300]{Computing methodologies~Neural networks}



\keywords{Multimodal Fusion; Pointing; Eye Gaze; Object Referencing; Personalized Models; Machine Learning; Deep Learning}


\maketitle

\section{Introduction}

As vehicles move towards fully autonomous driving, driving assistance functionalities are exponentially increasing, such as parking assistance, blind-spot detection, and cruise control. These features make the driving experience much easier and more enjoyable, which allows car manufacturers to design and implement numerous other additional features such as detecting eyes off-road using gaze monitoring, or enhancing secondary tasks while driving, such as music and infotainment control using mid-air hand gestures and voice commands. All these driving- and non-driving-related features require numerous sensors to work, such as external and internal cameras, gaze trackers, hand and finger detectors, and motion sensors. This enables researchers to design more novel features that could make use of the full range of available sensors, such as automatic music control using emotion recognition, and referencing objects (e.g., landmarks, restaurants, and buildings) outside the car. The latter feature is the focus of this paper.

\begin{figure}[!t]
	\begin{center}
		\includegraphics[width=0.9\linewidth]{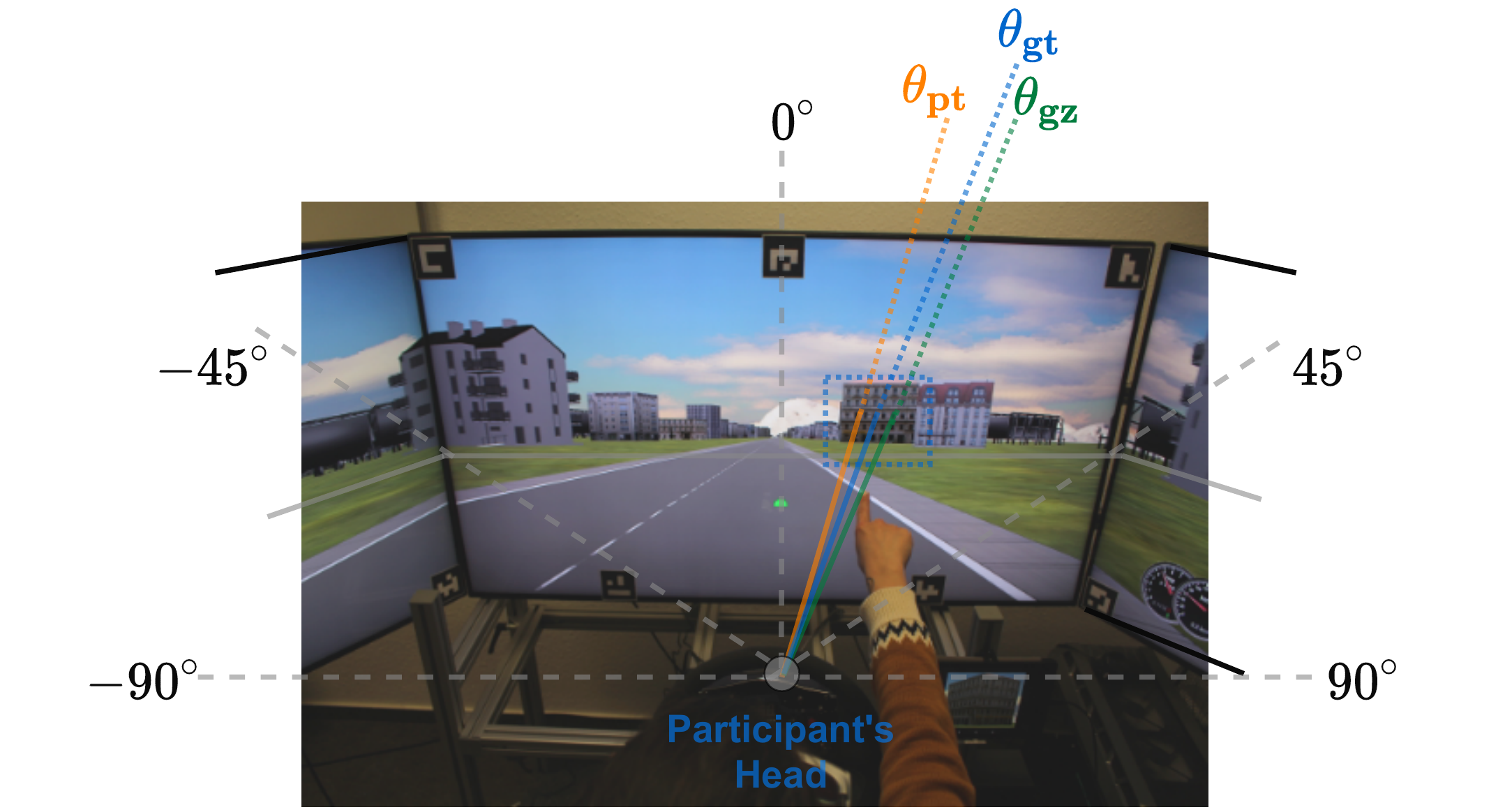}
	\end{center}
	\caption{The Referencing Task. $\theta_{pt}$ represents the horizontal pointing angle, $\theta_{gz}$ represents the horizontal gaze angle and $\theta_{gt}$ represents the center of the referenced object.}
	\label{fig:implementationarchitecture}
\end{figure}

Selecting and referencing objects using speech, gaze, hand, and pointing gestures have been attempted in multiple domains and situations such as human-robot interaction, Industry 4.0, and in-vehicle and outside-the-vehicle interaction. Only recently did car manufacturers start implementing gesture-based in-vehicle object selection, such as the \textit{BMW Natural User Interface}~\cite{2019BMW2019} using pointing, and outside-the-vehicle object referencing, such as the \textit{Mercedes-Benz User Experience (MBUX) Travel Knowledge Feature}~\cite{2021Mercedes-Benz2021} using gaze. However, these features focus on single-modality interaction or limited multimodal approach where the second modality (usually speech) is used as event trigger only. This may deteriorate the referencing task performance during high-stress dynamic situations such as high-traffic driving or lane changing~\cite{Szalma2004,Barua2020,Brown2020}.
While multimodal interaction is more natural and provides superior performance compared to single-modality approaches, it introduces inherent complexity and possesses multiple challenges, such as fusion techniques (e.g., Early versus Late Fusion~\cite{Snoek2005,Bianco2015,Seeland2017}), translation, co-learning, representation, and alignment~\cite{baltruvsaitis2018multimodal,atrey2010multimodal}. This has limited its use in previous systems, especially in complex dual-task situations where several modalities can be used in both tasks at the same time, such as infotainment interaction while driving. 

In this paper, we explore several fusion strategies to overcome these challenges during the referencing task. 
Specifically, we focus on machine-learning-based approaches for personalized referencing of outside-the-vehicle objects using gaze and pointing multimodal interaction.

\section{Related Work}

One-year-old infants learn to point at things and use hand gestures even before learning how to speak; hence, gestures are the most natural form of communication used by human beings~\cite{takemura2018neural}. However, gesturing is not exclusive to hand movements; it can be done using face motion or even simple gazing. Therefore, researchers have attempted various approaches for controlling objects using symbolic hand gestures~\cite{Zhao2019implGestureInfotain,Ye2019,Haria2017,Singh2015,Rempel2014}, deictic (pointing) hand gestures~\cite{nickel20043d,jing2013human,kehl2004real}, eye gaze~\cite{vidal2012detection,vidal2013pursuits,cheng2018smooth,Zhang2017Look,poitschke2011gaze,rayner2009eye,land2009looking,Moniri_2018,kang2015you}, and facial expressions~\cite{Abdat2011,GhorbandaeiPour2018,Singh2020}. 
Specifically for the automotive domain, in-vehicle interaction has been attempted using hand gestures~\cite{Ahmad2018a,Fariman2016,Roider2017,feld2016combine}, eye gaze~\cite{poitschke2011gaze}, and facial expressions~\cite{sezgin2009multimodal}, while outside-the-vehicle interaction has been attempted using pointing gestures~\cite{rumelin2013free,fujimura2013driver}, eye gaze~\cite{kang2015you}, and head pose~\cite{kim2014identification,misu2014situated}. Although most of the previous methods focus on single-modality approaches while using a button or voice commands as event triggers, more recent work focused on multimodal fusion approaches to enhance performance. Roider et al.~\cite{roider2018see} used a simple rule-based multimodal fusion approach to enhance pointing gesture performance for in-vehicle object selection while driving by including passive gaze tracking. They were able to enhance the overall accuracy, but only when the confidence in the pointing gesture accuracy was low. When the confidence was high, their fusion approach would have a negative effect and lead to worse results. 
Similarly, Aftab et al.~\cite{Aftab2020} used a more sophisticated deep-learning-based multimodal fusion approach combining gaze, pointing and head pose for the same task, but with more objects to select. 

While previous approaches, where both source and target are stationary, show great promise for in-vehicle interaction, they are not directly applicable to an outside-the-vehicle referencing task. This is because in this case, the source and the target are in constant relative motion. Moniri et al.~\cite{Moniri2012a} studied multimodal interaction in referencing objects outside the vehicle for passenger seat users using eye gaze, head pose, and pointing gestures and proposed an algorithm for the selection process. Similarly, Gomaa et al.~\cite{Gomaa2020} investigated a multimodal referencing approach using gaze and pointing gestures while driving and highlighted personal differences in behavior among drivers. They proposed a rule-based approach for fusion that exploits the differences in users' behavior by modality switching between gaze and pointing based on a specific user or situation. While both these methods align with the multimodal interaction paradigm of this paper, they do not strictly fuse the modalities together, but rather decide on which modality to use based on performance. They also do not consider time series analysis in their approaches, but rather treat referencing as a single-frame event.

Unlike these previous methods, this paper focuses on utilizing both gaze and pointing modalities through explicit autonomous fusion methods. Our contributions can be summarized as follows: 1) we propose a machine-learning-based fusion approach and compare it with the rule-based fusion approaches used in previous methods; 2) we compare an early fusion approach with a late fusion one; 3) we compare single frame analysis with time series analysis; 4) we utilize individuals' different referencing behavior for a personalized fusion approach that outperforms a non-personalized one.

\section{Design and Participants}

\begin{figure}[!t]
	\begin{center}
		\includegraphics[width=0.9\linewidth]{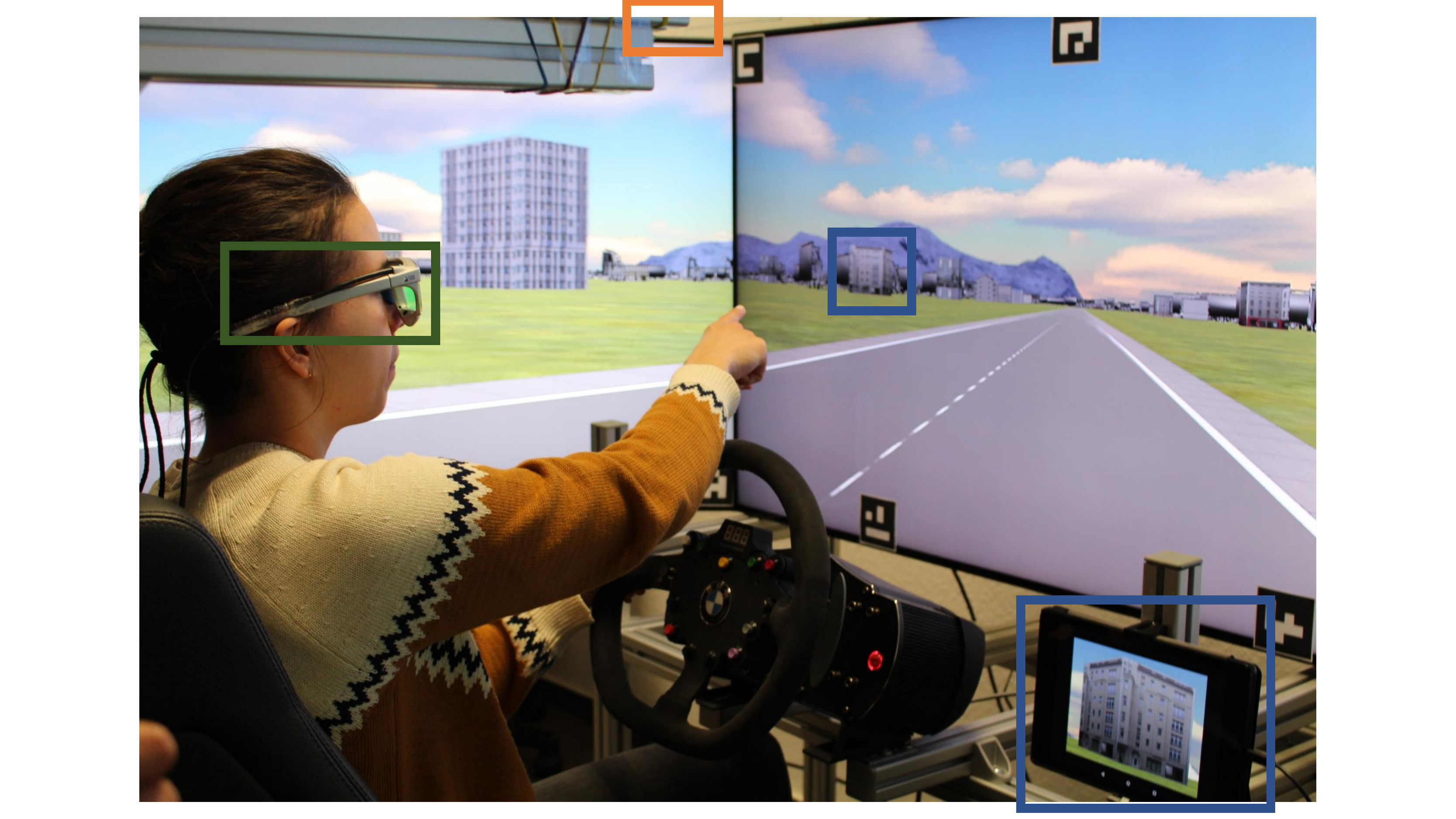}
	\end{center}
	\caption{Our driving simulator setup overview. The pointing tracker location is highlighted in orange, gaze tracking glasses are highlighted in green, and PoI is seen in the simulator and as a notification on the tablet highlighted in blue.}
	\label{fig:setup2}
\end{figure}

We designed a within-subject counterbalanced experiment using the same setup (as seen in~\autoref{fig:setup2}) as Gomaa et al.~\cite{Gomaa2020} to collect data for the referencing task. We used OpenDS software~\cite{math2013opends} with a steering wheel, pedals, and speakers for the driving simulation; a state-of-the-art non-commercial hand tracking camera prototype specially designed for in-vehicle control for pointing tracking; and SMI Eye Tracking Glasses\footnote{\url{https://imotions.com/hardware/smi-eye-tracking-glasses/}} for gaze tracking. For the driving task, we used  the same 40-minute five-conditions star-shaped driving route with no traffic and a maximum driving speed of 60 km/hour as Gomaa et al.~\cite{Gomaa2020}. The five counterbalanced conditions were divided into two levels of environment density (i.e., number of distractors around the Points of Interest (PoI)), two levels of distance from the road (i.e., near versus far PoI), and one autonomous driving condition. PoIs (e.g., Buildings) to reference are randomly located on the left and right of the road for all conditions and signaled to the driver in real-time. Drivers completed 24 referencing tasks per condition with a total of 124 gestures. These conditions were selected to implicitly increase the external validity of the experiment and the robustness of the learning models, however, there were not explicitly classified during model training.
Finally, the PoIs' shapes and colors were selected during a pre-study to subtly contrast the surroundings and a pilot study was conducted to tune the design parameters.

\begin{figure*}
	\begin{center}
		\includegraphics[width=0.9\linewidth]{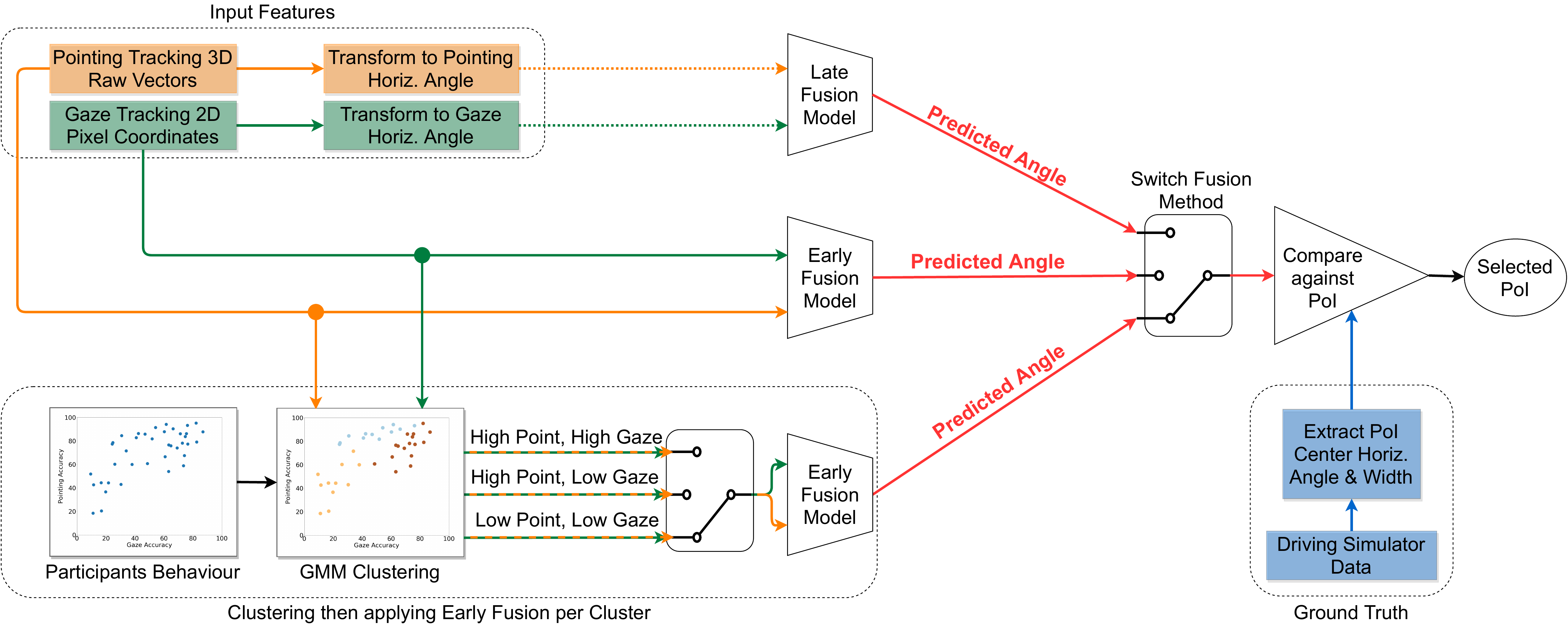}
	\end{center}
	\caption{\label{fig:architectureoverview}System architecture showing possible fusion approaches. Early fusion is an end-to-end machine learning approach while late fusion requires pre-processing the pointing and gaze modalities first, and extracting final horizontal angles for both using the approach in~\cite{Gomaa2020}.}
\end{figure*}

We recruited 45 participants for the study. Six were excluded: one due to motion sickness and failure to complete the entire route, another due to incorrectly executing the task, and the other four due to technical problems with the tracking devices. The remaining 39 participants (17 males) had a mean age of 26 years (SD=6). Thirty-two participants were right-handed while the rest were left-handed.

\section{Method}

\autoref{fig:architectureoverview} shows our proposed fusion approaches. We compared using late and early fusion on the pointing and gaze modalities to determine the referenced object. We also performed a hybrid learning approach that combines unsupervised and supervised learning to predict the PoI's center angle; we first cluster participants based on pointing and gaze behavior and then we train our early fusion model on each cluster to boost the fusion performance per user.

\subsection{Fusion Approach}

We consider a common 1D polar coordinate system with horizontal angular coordinates only (as seen in~\autoref{fig:implementationarchitecture}) to which both pointing and gaze tracking systems as well as the PoI ground truth (GT) are transformed (similar to~\cite{Gomaa2020,kang2015you}). Since this is a continuous output case, we need to solve a regression problem. We consider three fusion approaches depending on the input feature type as follows.

    \subsubsection{Late Fusion} 
    The first approach involves pre-processing of the data, in which we transform the pointing and gaze raw data using the method in~\cite{Gomaa2020}, then use these transformed horizontal angles as input features to our machine learning algorithm while using the PoI center angle as the target output (i.e., ground truth).
    \subsubsection{Early Fusion} 
    The second approach is an end-to-end solution where we use the raw 3D vectors outputted from the pointing tracker and the 1D x-axis pixel coordinates outputted from the gaze tracker as input features to the learning algorithm while the ground truth remains the same.
    \subsubsection{Hybrid Fusion} 
    The third approach is a hybrid approach that combines early fusion with a Gaussian Mixture Model (GMM)~\cite{Hand1989} clustering algorithm where the participants are divided into three clusters based on the accuracy of pointing and gaze (i.e., the distribution of pointing and gaze accuracy per participant as calculated in~\cite{Gomaa2020}). The three clusters are ``Participants with high pointing and high gaze accuracy'', ``Participants with high pointing and low gaze accuracy'', and ``Participants with low pointing and low gaze accuracy''; there were no participants with high gaze and low pointing accuracy. Then, pointing and gaze input features are considered for each cluster of participants separately when applying the early fusion method to predict the GT angle. This approach can be considered as a semi-personalized approach that exploits individual referencing performance on a group of users.

\subsection{Frame Selection Method}

We use the pointing modality as the main trigger for the referencing action, and thus we have multiple frame selection approaches to fuse it with the gaze modality. We define them as follows:

    \subsubsection{All Pointing, All Gaze}The first approach is to use all pointing and gaze frames as input modalities with the corresponding PoI center angle as GT. This produces, on average, five data points (i.e., frames) per PoI per user, resulting in an average of 620 data points per user that could be used for training or testing.
    \subsubsection{Mid Pointing, Mid Gaze} The second approach is to use the middle frame of pointing and the middle frame of gaze as input modalities with the corresponding PoI center angle as GT. This approach produces one data point per PoI per user, resulting in 124 data points per user that could be used for training or testing.
    \subsubsection{Mid Pointing, First Gaze} The third and final approach is to use the middle frame of pointing as pointing modality input with the corresponding PoI center angle as GT. However, we will take the gaze modality frame at the time of the first pointing frame. The reason behind this approach is that there is evidence that drivers tend to glance at an object only at the very beginning of the pointing action then focus on the road~\cite{rumelin2013free,Gomaa2020}. This approach produces the same number of data points per user as the second approach. 
\begin{figure}[!t]
	\begin{center}
		\includegraphics[width=0.75\linewidth]{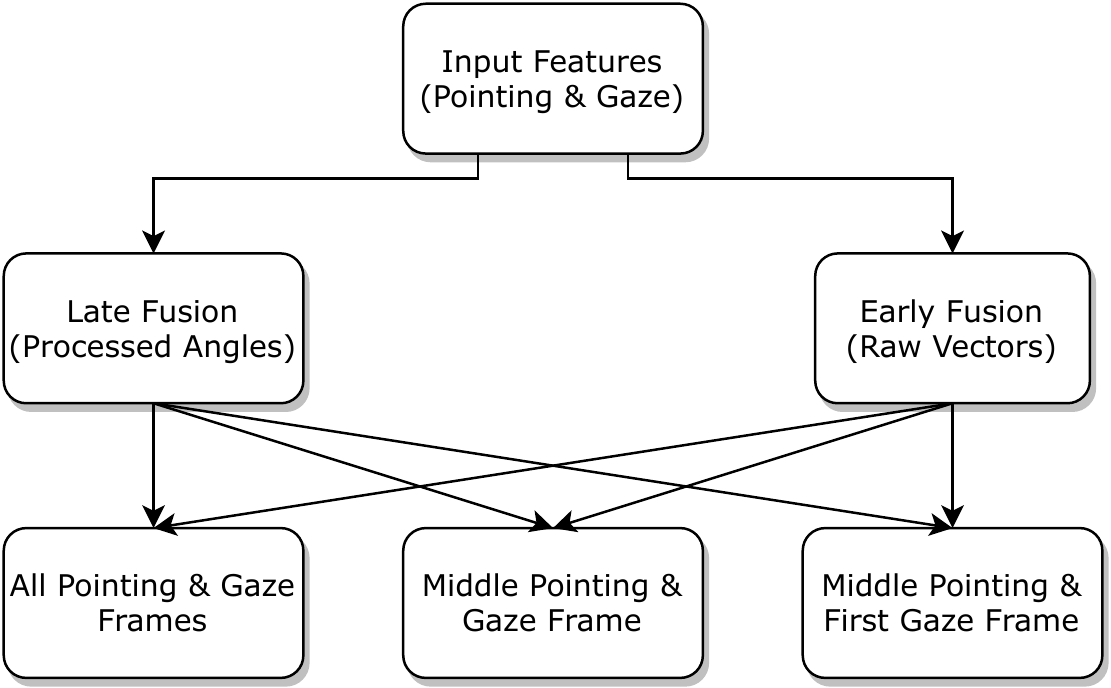}
	\end{center}
	\caption{Input features with possible different frame selection combinations for each fusion approach.}
	\label{fig:mapsvr}
\end{figure}

\autoref{fig:mapsvr} summarizes these different input feature approaches along with possible frame selection methods.

\begin{figure*}

	\begin{center}
		\includegraphics[width=0.7\linewidth]{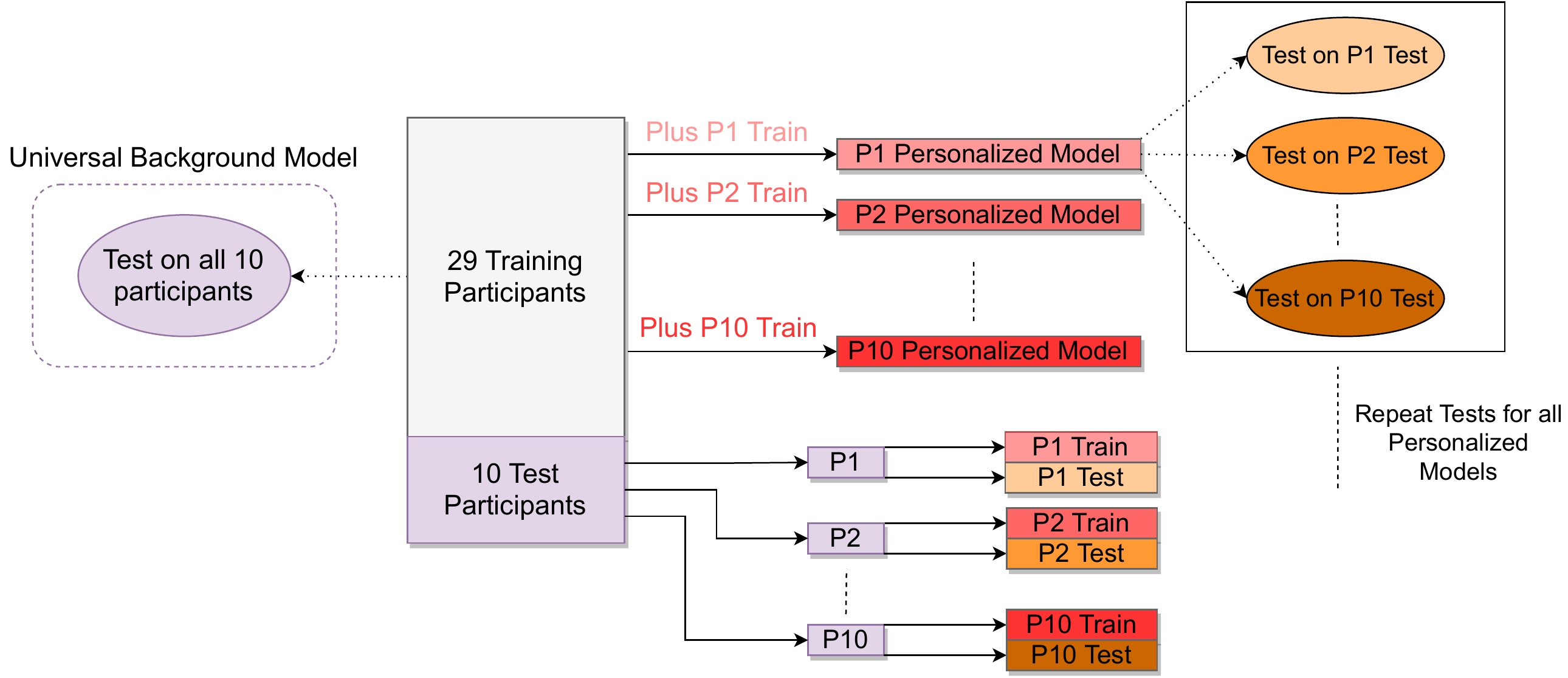}
	\end{center}
	\caption{\label{fig:traintestsplit}Dataset split approach for UBM and PM.}
\end{figure*}


\subsection{Dataset Split}

The entire dataset consisting of 39 participants is split into training, validation, and test sets using nested cross-validation~\cite{Browne2000} approach with 4-fold inner and outer loops. In addition, the test data is further split when training and testing personalized models (see~\autoref{fig:traintestsplit}). Note that all the splits are done on the participant level and not on the data point level. This way, no data from the same participant is used in training and testing for the same model, to ensure external validity and model generalization. Two models are implemented to satisfy generalization and personalization aspects as follows.

    \subsubsection{Universal Background Model (UBM)} This model follows the classic approach to machine learning: use training and validation data to create a prediction model with low generalization error reflected by its performance on the test data. A total of 29 participants are split into training and validation sets by the 4-fold inner loop of the nested cross-validation. It is then evaluated on the 10 participants selected for the test set by the 4-fold outer loop. 
    \subsubsection{Personalized Model (PM)} To train and test the personalization effect, we split the test data of each of the 10 participants into two halves, one that will be used as training data for personalization, and the other for personalization testing. Note that data points are shuffled before splitting for each participant. For each participant, we add the first half of the participant’s samples to the training samples from the UBM (that corresponds to 29 participants) and we retrain a Personalized Model (PM) that was adapted for this participant. This trained model is tested on the second half of the same participant’s data points as well as the data points of each of the other 9 test participants to verify the effect of personalization.

\subsection{Fusion Model}

For each of the fusion approaches and different frame selection approaches, we use a machine learning model for learning. Specifically, we implement a Support Vector machine for Regression model (SVR)~\cite{scholkopf1999shrinking}. We hypothesize that a deep learning model is not adequate in our use case due to the dataset's small size, however, we also compare different fusion models using deep learning approaches~\cite{bishop2006pattern,Goodfellow-et-al-2016} (e.g., Fully Connected Neural Network (FCNN), Convolutional Neural Network (CNN) and Long Short-Term Memory (LSTM)). Technical specifications for each model are as follows.

\subsubsection{Machine Learning Model: SVR}

Support Vector Machines (SVM)~\cite{cortes1995support} are a powerful machine learning algorithm based on the maximal margin and support vector classifier, which were first intended for linear binary classification. However, later extensions to the algorithms made it possible to train SVMs for multi-class classification and regression (SVR)~\cite{scholkopf1999shrinking} both linearly and non-linearly. In our setup, we use the python-based library \textbf{Scikit-Learn}~\cite{scikit-learn} for SVR implementation with a non-linear radial basis function (RBF) kernel and epsilon value of two, while the remaining parameters remain default values. These values for the hyperparameters were chosen using 4-fold cross-validation.

\subsubsection{Deep Learning Models: FCNN, CNN, and LSTM}

Deep learning or Artificial Neural Networks ~\cite{bishop2006pattern,Goodfellow-et-al-2016} are, in many regards, more powerful learning algorithms than traditional machine learning ones. However, they require much more training data to function optimally~\cite{Marcus2018DeepAppraisal}. We use the Python-based library \textbf{Keras}~\cite{chollet2015keras} for FCNN, CNN, and LSTM implementation. Due to the small size of the training dataset, we use a small number of layers for the neural network implementation to reduce the number of learnable parameters; the FCNN network consists of three hidden layers of size 32, 16, and 8 respectively, while non-linearity is introduced using the rectified linear unit (ReLU) activation function, and the 1D CNN network consists of one hidden convolutional layer of 64 filters and a kernel size of 3. Lastly, the LSTM network consists of one hidden layer having 50 hidden units. The output layer for all types of networks is a fully connected layer of output one and a linear activation function to work as regressors. Similar to the machine learning models, these architecture sizes (e.g., number of layers) and hyperparameters (e.g., filter and kernel sizes) were chosen using 4-fold cross-validation.

\subsection{Performance Metrics}

The most commonly used metric for regression problems is Mean Squared Error (MSE) or Root Mean Squared Error (RMSE), which would directly relate to the PoI center angle (i.e., GT). However, in our scenario, users might point (and gaze) at the edge of the PoI instead of the center, which still should be considered as correct referencing. For instance, a user might point at the edge of a wide building with a higher MSE than when pointing at the edge of a tall, narrow building. Therefore, we define another metric relating to the PoI width for performance comparison along with the RMSE.

\subsubsection{Root Mean Square Error (RMSE)}

\begin{figure}[!b]
	\begin{center}
		\includegraphics[width=\linewidth]{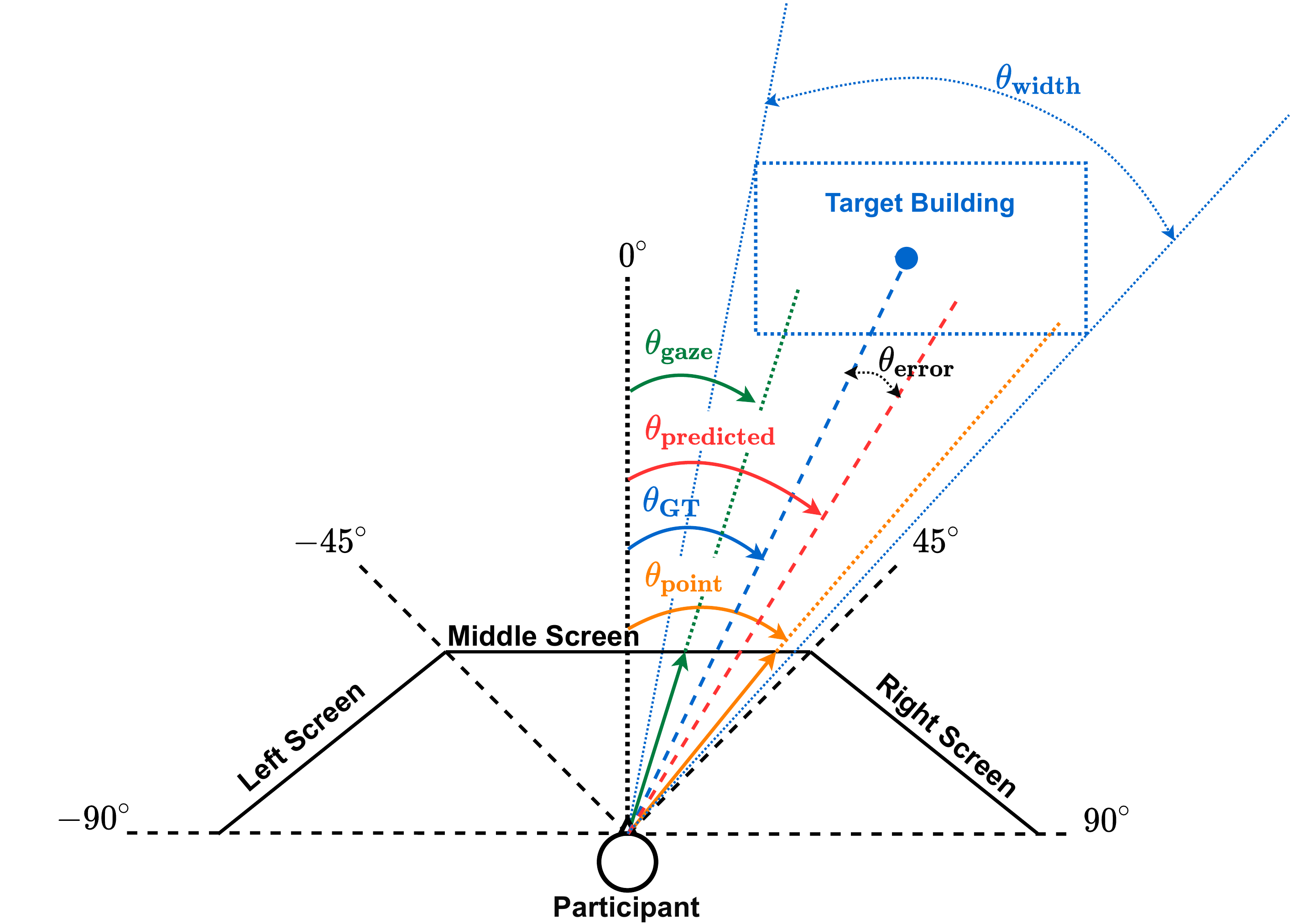}
	\end{center}
	\caption{An illustration showing example pointing and gaze vectors and their respective angles. It also shows the predicted angle out of these vectors and how it compares to the ground truth angle.}
	\label{fig:anglepredictionthetaerror}
\end{figure}

Root Mean Square Error (RMSE) relates directly to the output angle. ~\autoref{fig:anglepredictionthetaerror} further illustrates this RMSE metric. The machine learning model attempts to predict a fused angle of pointing and gaze ($\theta_{predicted}$) as close to the PoI center (i.e., ground truth) as possible. The difference between the predicted angle and the ground truth angle ($\theta_{error}$) is calculated for each data point and squared; then, the mean of all $N$ square errors is calculated, which results in the MSE. Finally, the RMSE is simply the square root of the MSE, which produces an average estimation of prediction error in degrees as seen in~\autoref{eqn:RMSE}.

\begin{equation}
\label{eqn:RMSE}
\begin{split}
\textit{RMSE} = \sqrt{\frac{1}{N}\sum_{i=1}^{N}{\theta_{error_i}^2}} = \sqrt{\frac{1}{N}\sum_{i=1}^{N}{(\theta_{prediction_i}-\theta_{GT_i})^2}}
\end{split}
\end{equation}

For example, if the resultant MSE of our prediction is 225, it means that the average error of the model prediction (i.e., RMSE) is 15 degrees away from the center of the PoI, so if we assume that we have a PoI's angular width of 40 degrees in visual space, it means that our model already predicts a referencing angle completely within the PoI. Note that a limitation of the RMSE metric is that it does not differentiate between positive and negative offset since it is squared during the MSE calculation. So, this 15-degree average error could be to the left or the right of the PoI center.

\subsubsection{Mean Relative Distance-agnostic Error (MRDE)}

The previous RMSE metric returns an average absolute error in degrees with which we, in theory, could assess our model performance. However, when interpreting this error, caution must be taken, since larger error values have a more severe impact for far PoIs versus near ones as far objects have much lower angular width ($\theta_{width}$) compared to closer objects. Thus, a relative metric would better describe the performance in comparison to an absolute one.
Consequently, we introduced the Relative Distance-agnostic Error (RDE) metric, which divides the difference between the predicted angle and the ground truth angle ($\theta_{error}$) by half of the angular width ($\frac{\theta_{width}}{2}$) of the PoI on the screen at this time instance (since the vehicle is moving, the angular width changes with time). The resulting value then describes the error in multiples of half the PoI's width at each time instance. Thus, the MRDE is the average of the RDE for all $N$ time instances (i.e., data points) as seen in~\autoref{eqn:MRDE}. Consequently, values less than one denote that the pointing vector lies within the PoI, while values larger than one denote that the referencing vector is outside the PoI’s boundaries.

\begin{equation}
\label{eqn:MRDE}
\begin{split}
\textit{MRDE} = \frac{1}{N}\sum_{i=1}^{N}{\frac{\theta_{error_i}}{0.5 \cdot \theta_{width_i}}} = \frac{1}{N}\sum_{i=1}^{N}{\frac{\theta_{prediction_i}-\theta_{GT_i}}{0.5 \cdot \theta_{width_i}}} 
\end{split}
\end{equation}

While this metric allows a better interpretation of the severity of the referencing error in this study, it still does not yield a perfect absolute rating, since it now depends on the size of PoIs used in this study. For the task at hand, however, we consider this sufficient since PoI sizes did not vary greatly across the experiment.

\section{Results and Discussion}

Results are divided into three sections. Section~\ref{sec:ubm_results} shows and discusses the UBM results and the selection of the best fusion approach. We implement the UBM using all fusion approaches, frame selection methods, and fusion models to determine the optimum fusion combination that is used in the personalization approach. Section~\ref{sec:pm_results} discusses the PM results and the selection of additional hyperparameters that are concerned with personalization only. Section~\ref{sec:bahavior_results} discusses some further investigations on the pointing behavior quantitatively. Additionally, we illustrate the effect of multimodal fusion in comparison to single-modality approaches in all sections. 

\subsection{Universal Background Model (UBM) Results}
\label{sec:ubm_results}

In this section, we first show the results of using a machine learning model, specifically SVR. Then, we compare against the FCNN and hybrid models. Lastly, we illustrate the effect of including data points in chronological order in the analysis (i.e., time series analysis) using CNN and LSTM.

\begin{figure}[!b]
     \centering
     \begin{subfigure}{\linewidth}
         \centering
         \includegraphics[width=\textwidth]{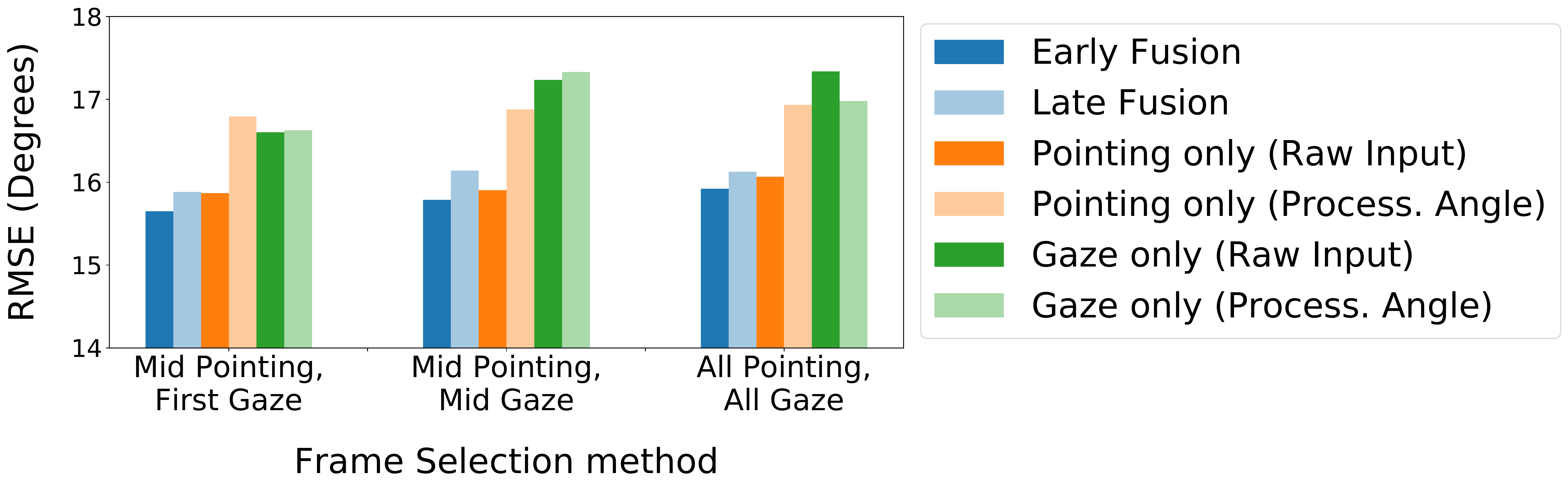}
         \caption{RMSE}
         \label{fig:ubm_earlyvslate_mse}
     \end{subfigure}
    \par\bigskip
     \begin{subfigure}{\linewidth}
         \centering
         \includegraphics[width=\textwidth]{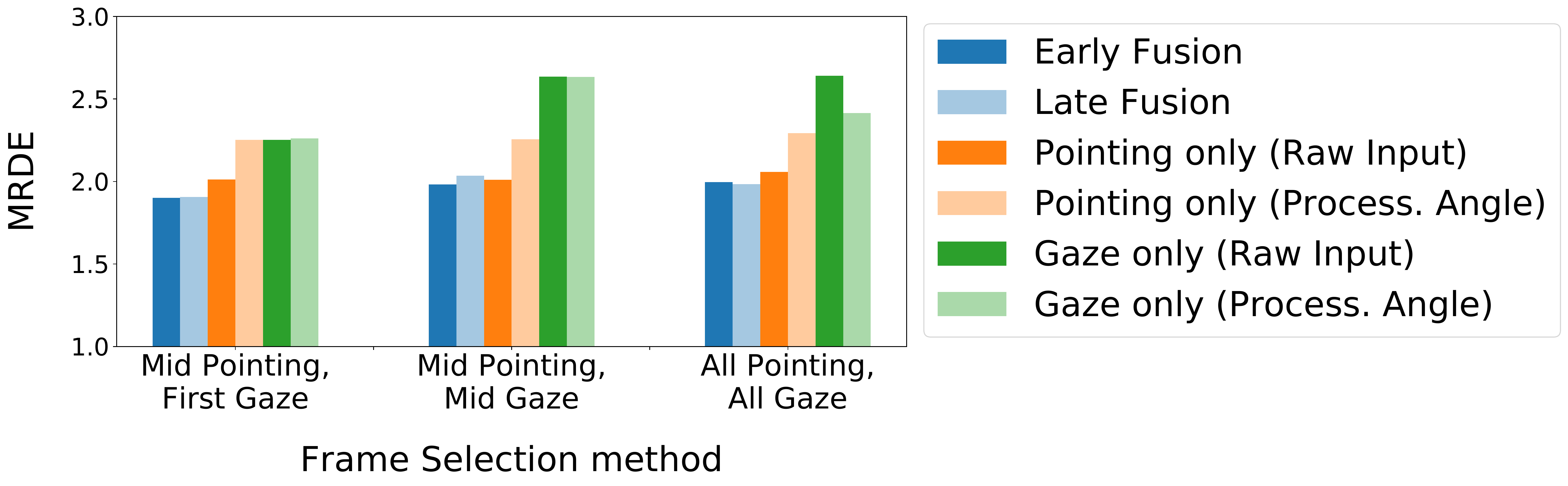}
         \caption{MRDE}
         \label{fig:ubm_earlyvslate_mrde}
     \end{subfigure}
     \caption{RMSE and MRDE results comparing UBM's Early versus Late Fusion for the multimodal interaction against prediction using single modality pointing and gaze for all frame selection approaches.}
\end{figure}

\subsubsection{UBM's Early vs. Late Fusion using SVR}

\autoref{fig:ubm_earlyvslate_mse} and~\autoref{fig:ubm_earlyvslate_mrde} show the comparison between early and late fusion approaches, using the RMSE and MRDE metrics respectively, against the single-modality approach, where lower values mean better performance as discussed earlier. Early fusion is superior to late fusion for RMSE values with a relative average of 3.01\%, while MRDE values are almost the same. Moreover, this early fusion approach outperforms pointing and gaze single-modality approaches by 2.82\% and 12.55\% respectively for RMSE and by 5.79\% and 18.55\% respectively for MRDE. 
Pointing modality is more accurate than gaze modality by 11.02\% average enhancement in RMSE and MRDE, and using raw pointing data outperforms pre-processing the pointing modality using the method in~\cite{Gomaa2020} by 12.01\% average enhancement in RMSE and MRDE. These results agree with previous work in multimodal interaction~\cite{Gomaa2020,Roider2017,roider2018see,Aftab2020} showing that while human interaction is multimodal in general, one modality would be dominant while the other modalities would be less relevant and their effect less significant in the referencing process.

The figures also illustrate the differences between possible pointing and gaze frame selection methods: while the performance difference is not statistically significant for any frame selection method over another, selecting the middle pointing frame and the first gaze frame still outperforms the other frame selection methods by an average of 2.65\% and 4.64\% for RMSE and MRDE respectively. This result also agrees with previous work~\cite{Gomaa2020,rumelin2013free} where participants usually glance at PoIs only at the beginning of the referencing task, while keeping their eyes on the road and pointing at the PoI during the remaining time of the referencing action.

\begin{figure}[!b]
     \centering
     \begin{subfigure}{\linewidth}
         \centering
         \includegraphics[width=\textwidth]{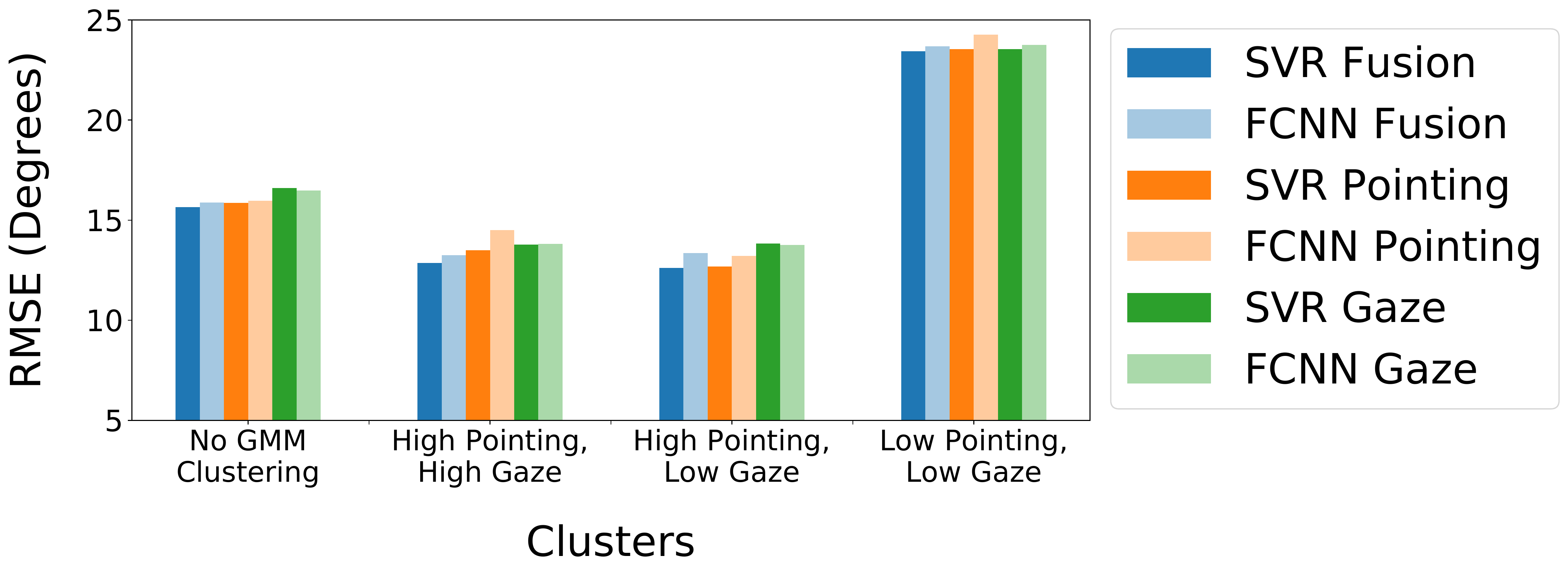}
         \caption{RMSE}
         \label{fig:ubm_svrvsfcnn_mse}
     \end{subfigure}
     \par\bigskip
     \begin{subfigure}{\linewidth}
         \centering
         \includegraphics[width=\textwidth]{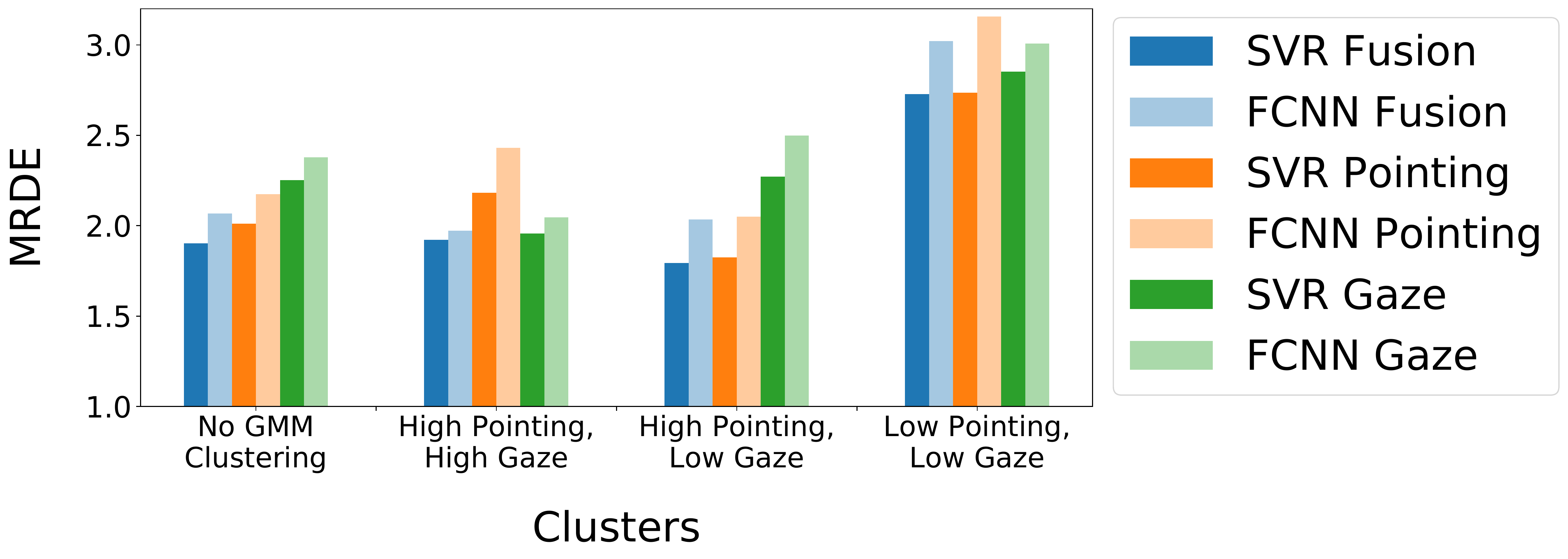}
         \caption{MRDE}
         \label{fig:ubm_svrvsfcnn_mrde}
     \end{subfigure}
     \caption{RMSE and MRDE results comparing UBM's Early Fusion versus Hybrid Early fusion approach when using the SVR algorithm or FCNN against single modalities. The results are calculated using the ``middle pointing and first gaze'' frame selection method.}
\end{figure}

\subsubsection{Comparison against FCNN and Hybrid Models}

Having chosen the ``middle pointing and first gaze'' frame selection approach as an optimum approach,~\autoref{fig:ubm_svrvsfcnn_mse} and~\autoref{fig:ubm_svrvsfcnn_mrde} show the comparison between the SVR model and the FCNN model. Early fusion using the SVR model is superior to using the FCNN model for both RMSE and MRDE values with a relative average of 2.91\% and 8.75\% respectively. Additionally, both pointing and gaze single-modality prediction using SVR is on average more accurate than using FCNN for both RMSE and MRDE by 4.57\% and 13.14\% respectively. Since the entire training dataset consists of only 3596 data points (i.e., 29 participants with 124 data points per participant), we attribute the superiority of SVR over FCNN to the small dataset size in our use case. This is because deep neural networks generally require more data (even for the shallow FCNN used), but this remains to be tested in the future.
Finally, the figures show a comparison between early fusion and hybrid fusion approaches. Fusion clusters (including high pointing accuracy, high gaze accuracy, or both) outperform the ``no GMM clustering'' approach for RMSE and MRDE by an average of 33.71\% and 3.38\% respectively, while the cluster where both modalities performed poorly (i.e., low pointing and low gaze accuracy) is degraded by 49.80\% and 43.45\% respectively. Moreover, for all clustering and no clustering cases, fusion models achieve better performance than single-modality ones. This means that the fusion model works better than the single best modality and does not get degraded by the other modality, unlike the rule-based fusion model in~\cite{roider2018see} that degrades with high pointing accuracy.

\begin{figure}[!b]
     \centering
     \begin{subfigure}{\linewidth}
         \centering
         \includegraphics[width=0.85\textwidth]{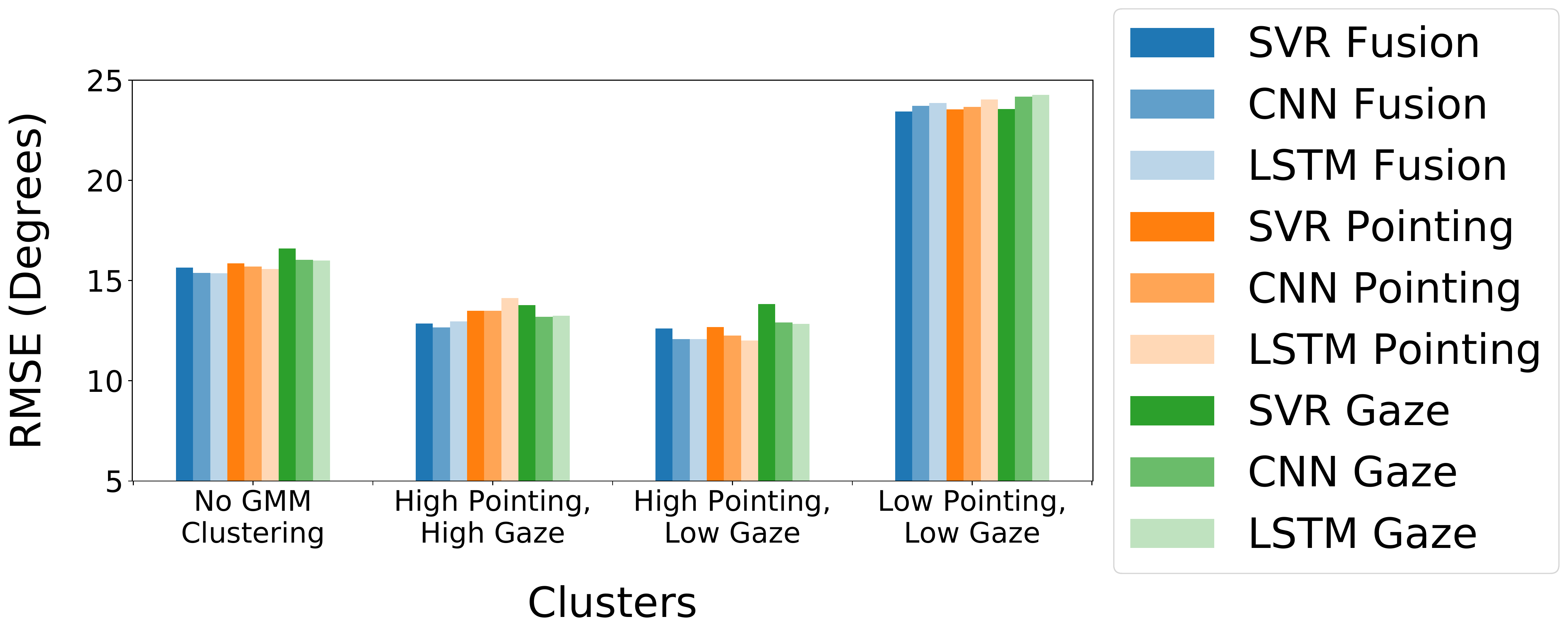}
         \caption{RMSE}
         \label{fig:ubm_svrvscnnvslstm_mse}
     \end{subfigure}
     \par\bigskip
     \begin{subfigure}{\linewidth}
         \centering
         \includegraphics[width=0.85\textwidth]{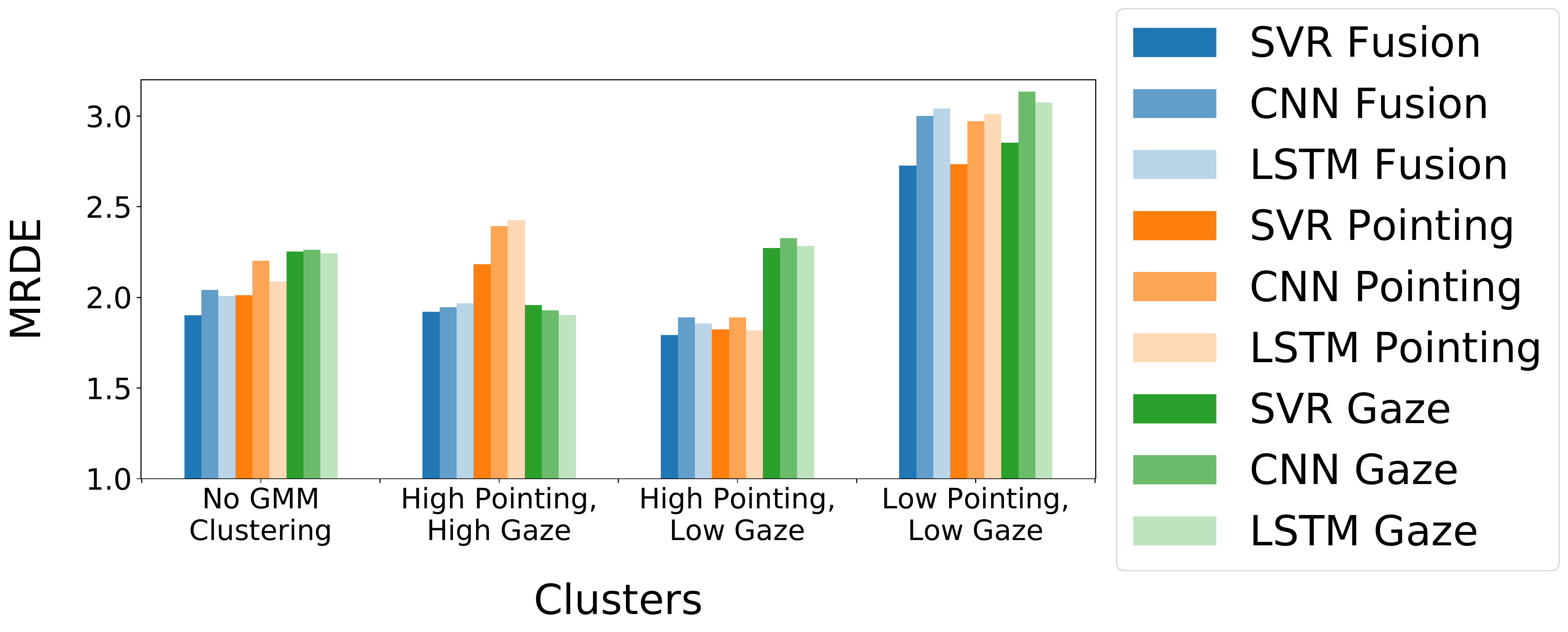}
         \caption{MRDE}
         \label{fig:ubm_svrvscnnvslstm_mrde}
     \end{subfigure}
     \caption{RMSE and MRDE results comparing UBM's Early Fusion versus the Hybrid Early Fusion approach when using the SVR algorithm, CNN or LSTM against single modalities. The results are calculated using the ``middle pointing and first gaze'' frame selection method.}
\end{figure}

\subsubsection{Time Series Analysis using CNN and LSTM}

Since both the CNN and LSTM algorithms require temporal information, the ``all pointing and all gaze'' frame selection method is used in analyses for both CNN and LSTM models as well as the previous SVR model for comparison.~\autoref{fig:ubm_svrvscnnvslstm_mse} and~\autoref{fig:ubm_svrvscnnvslstm_mrde} show the results for these analyses while comparing the early fusion model to the hybrid fusion model as in the previous section. From the figures, it can be seen that while the temporal approach methods (i.e., CNN and LSTM) are almost equivalent for all cases in both RMSE and MRDE metrics, they are quite different from the non-temporal one (i.e., SVR). Although temporal models are slightly superior to the non-temporal model for RMSE values (\autoref{fig:ubm_svrvscnnvslstm_mse}) with an average of 3.26\%, they are inferior in terms of MRDE values (\autoref{fig:ubm_svrvscnnvslstm_mrde}) by an average of 7.33\%, which suggests that a non-temporal approach is still the optimum approach. We hypothesize that the reason for that could also be the small dataset size, or it might be due to the fact that the temporal information for this task is too noisy or redundant to be useful for the fusion.

\subsection{Personalized Model (PM) Results}
\label{sec:pm_results}

\begin{figure}[!b]
     \centering
     \begin{subfigure}{\linewidth}
         \centering
         \includegraphics[width=0.75\textwidth]{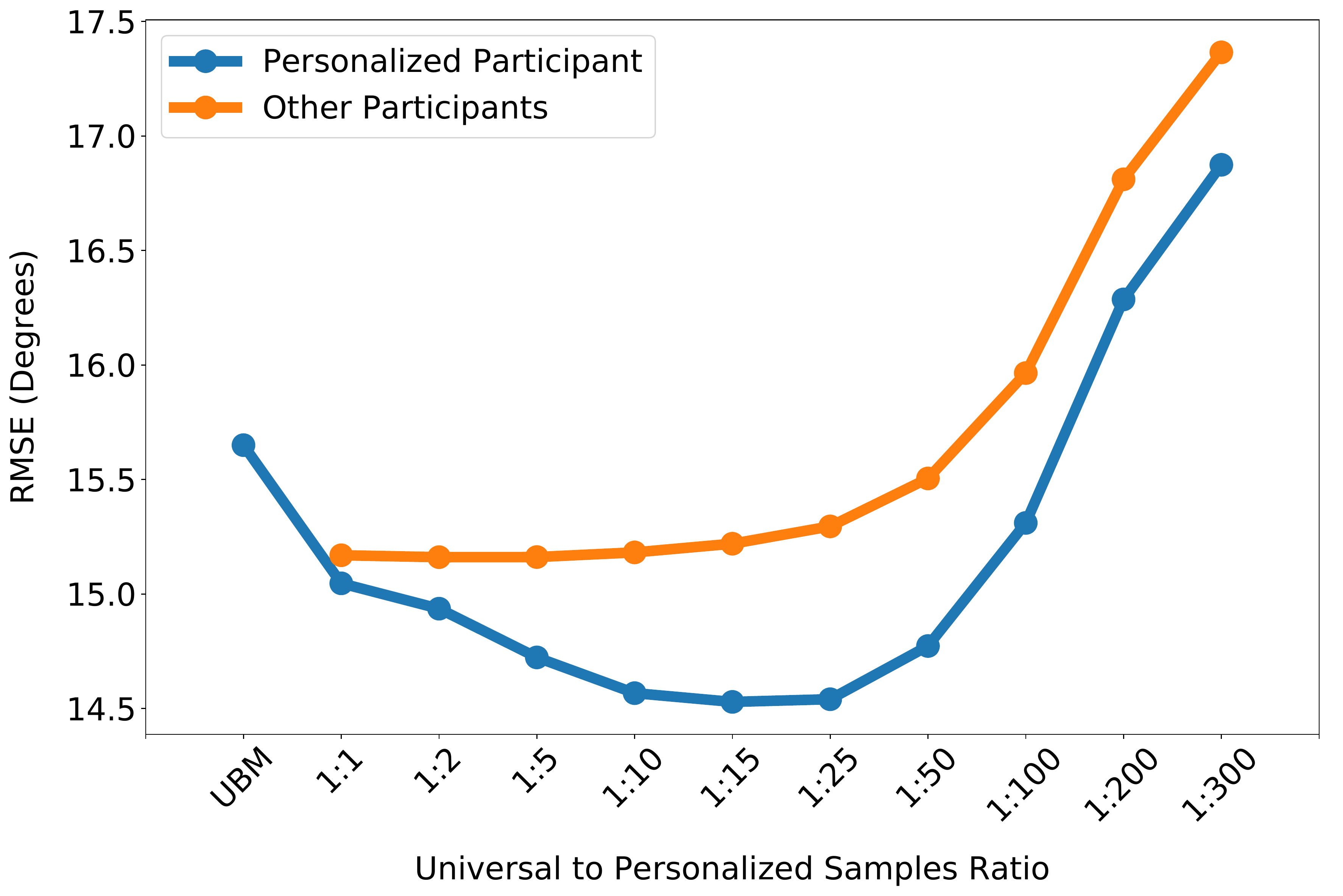}
         \caption{RMSE}
         \label{fig:PM_sample_weight_RMSE}
     \end{subfigure}
     \par\bigskip
     \begin{subfigure}{\linewidth}
         \centering
         \includegraphics[width=0.75\textwidth]{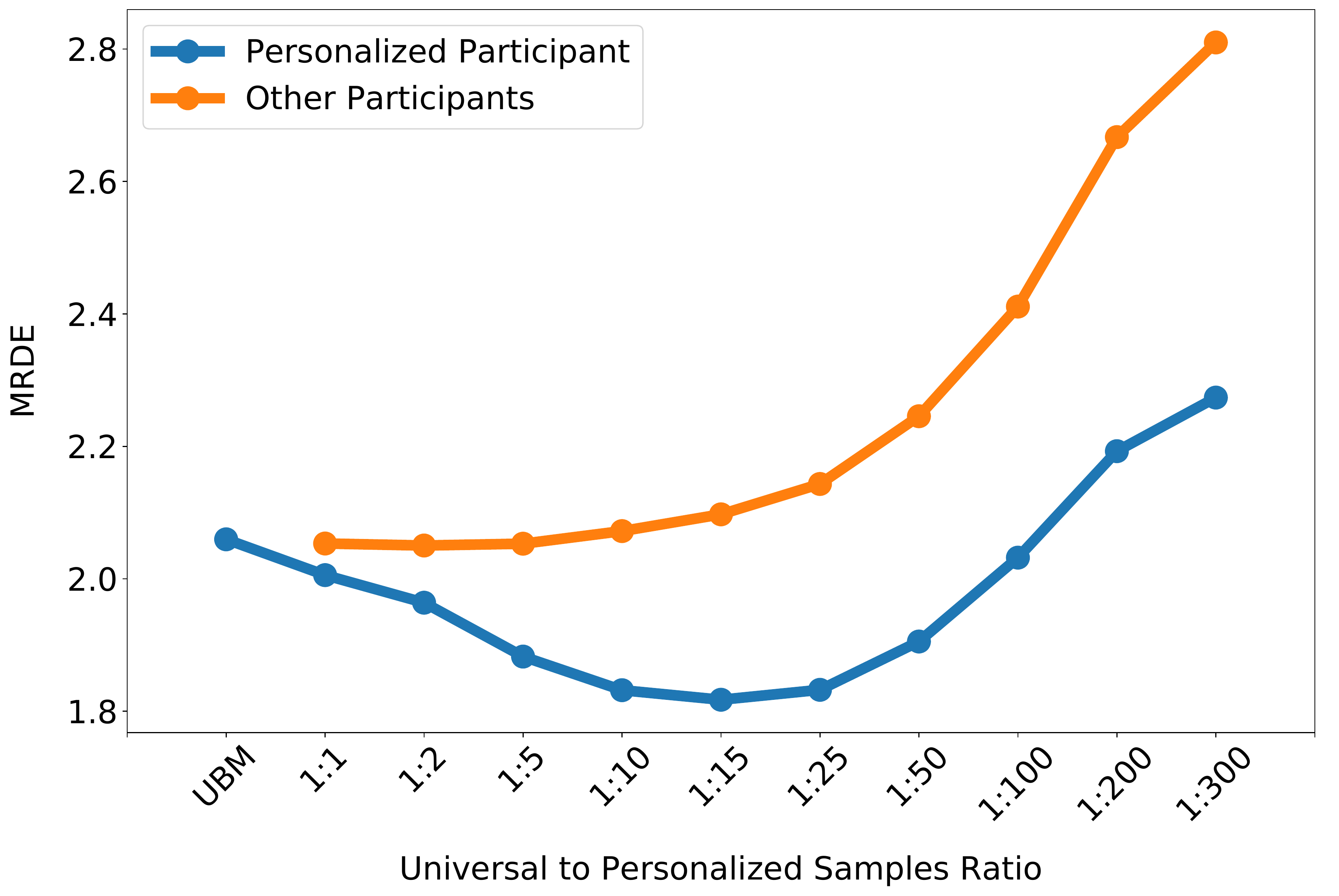}
         \caption{MRDE}
         \label{fig:PM_sample_weight_mrde}
     \end{subfigure}
     \caption{Fusion RMSE and MRDE results comparing PM with different sample weights when adding the personalized participant data to the trained UBM model and retraining for individual effect. The PM is also compared to the UBM model without any personalization. The ``Personalized Participant'' RMSE values in the graph are the average of the test results for all PMs for each participant. The ``Other Participants'' values are the average of the test results of the other non-personalized participants using the PM. All results are calculated using the ``middle pointing and first gaze'' frame selection method and the SVR algorithm.}
\end{figure}

As seen in~\autoref{fig:traintestsplit}, the PM is implemented by concatenating the overall training data points (i.e., all of the 29 participants' data points) with the training data points of each of the 10 test participants separately, then retraining this new dataset with a fusion model. However, when retraining using SVR, non-support-vector data points can be disregarded, as only the support vectors returned from the UBM training contribute to the algorithm. These support vectors are usually around 600 data points, which are combined with 62 data points (i.e., half of the participants' data points) during personalization. This low ratio (around 10\%) of individual participants' data points' contribution to the retraining process significantly reduces the personalization effect. 
Therefore, we introduced a sampling weight hyperparameter that would increase the importance of newly added data points when retraining by putting more emphasis on these data points by the SVR regressor.

\autoref{fig:PM_sample_weight_RMSE} and~\autoref{fig:PM_sample_weight_mrde} show the RMSE and MRDE results for the PM with the effect of this sampling weight, respectively, and they compare the PM to the UBM. It can be seen that the PM is superior to the UBM even if the personalized participant training data points (i.e., personalized samples) are weighted one-to-one with the other training participants' training data points (i.e., UBM samples); however, when giving more weight to the personalized participants' samples, the model adapts to the test data of these participants until we reach a point where the model starts overfitting and it does not perform well for this participant and the others. This point can be seen around a sampling weight ratio of one-to-fifteen in both~\autoref{fig:PM_sample_weight_RMSE} and~\autoref{fig:PM_sample_weight_mrde}, which shows a relative average enhancement for PM over UBM by 7.16\% and 11.76\% for RMSE and MRDE respectively. To further prove the personalization effect, it is tested against other participants that the model has not been personalized or adapted for. It can be seen from the same figures that, on average, PM works best for the personalized participant for all values of sample weight, with the largest gap around the one-to-fifteen optimum point. Further analysis on participants' level shows that at least 70\% of the participants conform to these average figures and their referencing performance enhances with personalization. 
In conclusion, while there could be other personalization techniques that would enhance the referencing task performance further, these results show that our suggested PM outperforms a generalized one-model-fits-all approach.   

\subsection{Referencing Behavior Analysis}
\label{sec:bahavior_results}

From~\autoref{eqn:MRDE}, we defined relative distance-agnostic error as the division of the difference between the predicted angle and the ground truth angle ($\theta_{error}$) by half of the angular width ($\frac{\theta_{width}}{2}$). However, it was noticed that at the middle frame of pointing, most of the buildings have very small angular widths ($\theta_{width}$). Additionally, it was noticed that participants’ average ground truth angle ($\theta_{GT}$) at the middle pointing frame was $\pm 9.8$ degrees (i.e., 9.8 degrees to the left or the right) and the ninety percent quantile of the ground truth angle was $\pm 17.3$ degrees. This means that most of the time participants did not wait for the building to get closer (a close building angle being around 35 to 45 degrees) and pointed rather quickly as soon as they identified it on the horizon. This is also confirmed by the average pointing time of 3.69 seconds reported in~\cite{Gomaa2020}, which shows that the users point quite quickly when the building appears, rather than waiting, even though the building remains in their view for approximately 20 seconds. 

To leverage from this observation, a slightly modified task was created by removing the pointing attempts whose ground truth exceeded $\pm 17.3$ degrees (i.e., the 90\% quantile threshold) from the dataset and analyzing only pointing at faraway buildings, since pointing at close buildings corresponds to only 10\% of this dataset and might not be representative. Then the PM's average RMSE and MRDE values are enhanced from 14.53 and 1.81 respectively (for the original task) to 5.45 and 0.544 respectively (for the modified far-pointing only task). This could be evidence that users attempt to point more accurately and gaze more precisely when referencing faraway objects to identify them, while they tend to be less accurate and more careless when pointing at near objects. In a real car environment, this observation is quite useful, since the surrounding world is captured using the car exterior camera and PoI angular width and distance can be easily identified.
Moreover, the data was further split based on the $\pm 9.8$ average pointing angle to compare the pointing behavior at very faraway buildings (below this average angle) and faraway buildings (above this average angle). The RMSE and MRDE values for both those cases were: 4.83 and 0.64 for the very far case, and 6.67 and 0.63 for the far case. This shows a correlation between the distance of the object from the user and the referencing behavior which supports the previously mentioned hypothesis that users point more accurately at more distant buildings as seen from the better RMSE value.

\section{Conclusion and Future Work}


Participants behave quite differently in multimodal interaction when referencing objects while driving. While previous work (such as~\cite{Gomaa2020}) showed pointing and gaze behavior and discussed the differences in users’ pointing behavior quantitatively and qualitatively towards modality fusion, they did not investigate the implementation of the fusion model itself. In this paper, we focus more on leveraging this information and the implementation of a multimodal fusion model using both pointing and gaze to enhance the referencing task. We also compare such generalized models to a more user-specific personalized one that adapts to these differences in user behavior.
Several machine learning algorithms have been trained and tested for the multimodal fusion approach and compared against the single-modality one. Also, different frame selection methods have been studied to assess the effect of time-series data on the performance when compared to single-frame methods. 

UBM results show that, on average, a multimodal fusion approach is more suitable than a single modality approach. They also show that a machine learning-based approach works better than deep learning, and time series analysis does not add much enhancement in performance compared to single-frame methods. However, this could be an artifact due to the dataset's small size, which we shall address in future work with longer experiment durations to collect more data.
As for the personalization effect, non-weighted PMs show a slight enhancement over the UBM, which could be attributed to the low contribution ratio of the participants' data points. 
On the other hand, weighted PMs show a large enhancement in the referencing task performance for that participant due to the emphasis on their data points (i.e., individual behavior).

Finally, the proposed fusion models, frame selection methods, and PM approaches enhance the referencing performance in comparison to both single-modality models and UBM. However, some aspects, like different pointing behavior based on the relative distance of the object to the user, are obscured through averaging over all data points, and this is not addressed in this study. In future work, these aspects require additional studies tailored towards this goal and further investigation to find a more concrete interpretation.

\begin{acks}
This work is partially funded by the German Ministry of Education and Research (Project CAMELOT, Grant Number: 01IW20008).
\end{acks}

\bibliographystyle{ACM-Reference-Format}
\balance
\bibliography{sample-base_not_linked}










\end{document}